# Data Descriptor

## Title
Multi-channel EEG recordings during a sustained-attention driving task

## Authors
Zehong Cao[1], Chun-Hsiang Chuang[1], Jung-Kai King[2], Chin-Teng Lin[1*]

## Affiliations
1. CI-BCI Lab, Centre for Artificial Intelligence, University of Technology Sydney, Sydney, Australia
2. Brain Research Centre, National Chiao Tung University, Hsinchu, Taiwan

*Corresponding author: Chin-Teng Lin (Chin-Teng.Lin@uts.edu.au)

## Abstract
We described driver behaviour and brain dynamics acquired from a 90-minute sustained-attention task in an immersive driving simulator. The data include 62 copies of 32-channel electroencephalography (EEG) data for 27 subjects that drove on a four-lane highway and were asked to keep the car cruising in the centre of the lane. Lane-departure events were randomly induced to make the car drift from the original cruising lane towards the left or right lane. A complete trial includes events with deviation onset, response onset, and response offset. The next trial, in which the subject has to drive back to the original cruising lane, occurs 5–10 seconds after finishing the current trial. We hope that this dataset will lead to the development of novel neural processing assays that can be used to index brain cortical dynamics and detect driving fatigue and drowsiness. This publicly available dataset is beneficial to the neuroscientific and brain-computer interface communities.

## Background & Summary

Driving safety has attracted public attention due to the increasing number of road traffic crashes. Risky driving behaviours, such as fatigue and drowsiness, increase drivers' risk of crashing, as fatigue suppresses driver performance, including awareness, recognition and directional control of the car [1]. In particular, high levels of fatigue and drowsiness diminish driver arousal and information processing abilities in unusual and emergency situations [2].

During a sustained-attention driving task, fatigue and drowsiness are reflected in driver behaviours and brain dynamics [3]. Furthermore, electroencephalogram (EEG) is the preferred method for human brain imaging when performing tasks involving natural movements in a real-world environment [4]. In 2003, we began conducting laboratory-based experiments collecting EEG data to investigate brain function associated with sustained attention on a safe driving task [5,6]. Our experiments have two distinct two goals: 1) neurocognitive performance: designing key signatures of how the neurocognitive state of the driver (e.g., physical and physiological) varies when faced with the sensory, perceptual and cognitive demands of a sustained-attention situation [7-9]; and 2) advanced computational approaches: investigating novel computational, statistical modelling and data visualisation techniques to extract signatures of neurocognitive performance, including novel analytic and algorithmic approaches for individually assessing drivers' neurocognitive state and performance [10,11].

In terms of the dataset, the experiment adopted an event-related lane-departure paradigm in a virtual-reality (VR) dynamic driving simulator to quantitatively measure brain EEG dynamics along with the fluctuation of behaviour performance. All of the participants were required to have a driver's licence, and none of them had a history of psychological disorders. The 32-channel EEG signals and vehicle position were recorded simultaneously, and all of the participants were instructed to sustain their attention in this driving experiment.

Several research studies on driving performance, including kinaesthetic effect, mind-wandering trends and the development of drowsiness prediction systems, have been conducted by our team using this EEG dataset. Specifically, to study EEG dynamics in response to kinaesthetic stimuli during driving, we used a VR-based driving simulator with a motion platform to produce somatic sensation similar to real-world situations [12]. For mind-wandering trends, we investigated brain dynamics and behavioural changes in individuals experiencing low perceptual demands during a sustained-attention task [13]. In terms of the drowsiness prediction system, we proposed a brain-computer interface-based approach using spectral dynamics to classify driver alertness and predict response times [14-18]. We determined the amount of cognitive state information that can be extracted from noninvasively recorded EEG data and the feasibility of online assessment and rectification of brain networks exhibiting characteristic dynamic patterns in response to cognitive challenges.

This data descriptor described a large EEG dataset in a sustained-attention driving task. It aims to help researchers reuse this dataset to understand the behavioural decision making of drivers under stress and cognitive fatigue in complex operational environments, such as car driving with kinaesthetic stimuli, which requires directly studying the interactions between brain, behavioural, sensory and performance dynamics based on their simultaneous measurement and joint analysis. We expect that this dataset could be used to explore principles and methods that can be used to design individualised real-time neuroergonomic systems to enhance the situational awareness and decision making of drivers under several forms of stress and cognitive fatigue, thereby improving total human-system performance. We believe this research will benefit the neuroscientific and brain-computer interface communities.

## Methods

### Participants

Twenty-seven voluntary participants (age: 22–28 years) who are students or staff of the National Chiao Tung University were recruited to participate in a 90-minute sustained-attention driving task at multiple times on the same or different days. In total, 62 EEG data were collected from these participants. Of note, the participants have normal or corrected-to-normal vision. In addition, none of the participants reported sleep deprivation in the preceding weeks, and none had a history of drug abuse according to the self-report. Every participant was required to have a normal work and rest cycle, get enough sleep (approximately 8 hours of sleep each night) and not stay up late (no later than 11:00 PM) for a week before the experiment. Additionally, the participants did not imbibe alcohol or caffeinated drinks or participate in strenuous exercise a day before the experiments. At the beginning of the experiment, a pre-test session was conducted to help participants understand the instructions and to confirm that none were afflicted with simulator-induced nausea. This study was performed in strict accordance with the recommendations in the Guide for Committee of Laboratory Care and Use of the National Chiao Tung University, Taiwan. The Institutional Review Board of the Veterans General Hospital, Taipei, Taiwan, approved the study. All of the participants were asked to read and sign an informed consent form before participating in the EEG experiments. The monetary compensation for one experimental session was approximately USD $20.

**Virtual-reality driving environment**

A VR driving environment with a dynamic driving simulator mounted on a six-degree-of-freedom Stewart motion platform was built to mirror reality behind the wheel. Six interactive highway driving scenes synchronised over local area networks were projected onto the screens at viewing angles of 0°, 42°, 84°, 180°, 276° and 318° to provide a nearly complete 360° visual field. The dimensions of the six directional scenes were 300 × 225 (width × height) cm, 290 × 225 cm, 260 × 195 cm, 520 × 195 cm, 260 × 195 cm, and 290 × 225 cm, respectively.

As shown in Figure 1-A and B, the experimental scenario involved a visually monotonous and unexciting night-time drive on a straight four-lane divided highway without other traffic. The distance from the left side to the right side of the road and the vehicle trajectory were quantised into values from 0–255, and the width of each lane was 60 units. The refresh rate of the scenario frame was set to emulate cruising at a speed of 100 km/hr. A real vehicle frame (Make: Ford; Model: Probe) (Figure 1-C) that included no unnecessary weight (such as an engine, wheels, and other components) was mounted on a six-degree-of-freedom Stewart motion platform (Figure 1-D). In addition, the driver's view of the VR driving environment driver was recorded and is shown in Figure 1-E.

**Experimental paradigm**

A event-related lane-departure paradigm [19] was implemented in the VR-based driving simulator using WorldToolKit (WTK) R9 Direct and Visual C++. The paradigm was designed to quantitatively measure the subject's reaction time to perturbations during a continuous driving task. The experimental paradigm simulated night-time driving on a four-lane highway, and the subject was asked to keep the car cruising in the centre of the lane. The simulation was designed to mimic a non-ideal road surface that caused the car to drift with equal probability to the right or left of the third lane. The driving task continued for 90 minutes without breaks. Drivers' activities were monitored from the scene control room via a surveillance video camera mounted on the dashboard. Lane-departure trials were collected from experimental data collected from 2005 to 2012 in National Chiao Tung University, Taiwan.

As shown in Figure 2-A, lane-departure events were randomly induced to make the car drift from the original cruising lane towards the left or right sides (deviation onset). Each participant was instructed to quickly compensate for this perturbation by steering the wheel (response onset) and to let the car drive back to the original cruising lane (response offset). To avoid the impacts of other factors during the task, participants only reacted to the lane-perturbation event by turning the steering wheel and did not have to control the accelerator or brakes pedals in this experiment. Each lane-departure event is defined as a "trial," including baseline period, deviation onset, response onset and response offset. EEG signals were recorded simultaneously (Figure 2-B). Additionally, the corresponding directions of turning the steering wheel are shown in Figure 2-C. Of note, the next trial occurs within a 5–10 second interval after finishing the current trial, in which the subject has to drive back to the centre line of the third car lane. If a participant fell asleep during the experiment, there was no feedback to alert him/her.

**Tutorial and code availability**

We hope readers like tutorial and code from figshare.com. To access these items, please go to the URL: https://doi.org/10.6084/m9.figshare.6427334.v2. Of them, a 59-page tutorial named "Tutorial Data Analysis for Multi-channel EEG Recordings during a Sustained-attention Driving Task.pdf" is provided for researchers to pre-process and analyse multi-channel EEG during a sustained-attention driving task. Furthermore, MATLAB codes named "Code-availability.zip" for EEG pre-processing and data analysis can also be found here.

## Data Records

**Data recording and storage**

During the experiment, the stimulus computer that generated the VR scene recoded the trajectories of the car as well as the events with time points in a "log" file. The stimulus computer also sent synchronised triggers (also recorded in the "log" file) to the Neuroscan EEG acquisition system. Concurrently, the Neuroscan system recoded EEG data with the time stamps of triggers in an "ev2" file. Because the number of time points in both recorded files were different, the first step was to integrate the two files into a new file with aligned event timing and behavioural data. The new event file was then imported by EEGLAB in MATLAB.

EEG signals were obtained using Scan SynAmps2 Express system (Compumedics Ltd., VIC, Australia). Recorded EEG signals were collected using a wired EEG cap with 32 Ag/AgCl electrodes, including 30 EEG electrodes and 2 reference electrodes (opposite lateral mastoids). The EEG electrodes were placed according to a modified international 10–20 system. The contact impedance between all electrodes and the skin was kept under 5 kΩ. The EEG recordings amplified by the Scan SynAmps2 Express system (Compumedics Ltd., VIC, Australia) were digitised at 500 Hz (resolution: 16 bits). Neuroscan's Scan 4.5 is the ultimate tool for data acquisition. The acquired raw data can be saved as .cnt files on the PC and server.

**EEG signals**

The raw files can be read using the EEGLAB toolbox in MATLAB. The uploaded files named with set suffixes contain all of the signals. After loading the files, the "EEG.data" variable includes 32 EEG signals and one vehicle position. The first 32 signals were from the Fp1, Fp2, F7, F3, Fz, F4, F8, FT7, FC3, FCZ, FC4, FT8, T3, C3, Cz, C4, T4,

TP7, CP3, CPz, CP4, TP8, A1, T5, P3, PZ, P4, T6, A2, O1, Oz and O2 electrodes. Two electrodes (A1 and A2) were references placed on the mastoid bones. The 33rd signal is vehicle position, which is used to describe the position of the simulated vehicle. Additionally, the types of events (see "EEG.event.type") in the dataset were classified into deviation onset (mark: 251 or 252), response onset (mark 253) and response offset (mark 254). As shown in Figure 3, we gave an example of behaviour performance (Figure 3-A) and EEG signals (Figure 3-B) with associated events.

## Technical Validation

### Behavioural validation

The EEG dataset was collected from 27 subjects with normal or corrected-to-normal vision. No subjects reported a history of psychiatric disorders, neurological disease or drug use disorders. All of the subjects were recruited university students and staff at the National Chiao Tung University, Taiwan. At the beginning of the experiment, each subject wore a suitable cap for recording the physiological data and were given 5 to 10 minutes to read the experimental instructions and complete the participant information sheet (questionnaire).

The EEG signals were recorded using Ag/AgCl electrodes attached to a 32-channel Quik-Cap (Compumedical NeuroScan). Thirty electrodes were arranged according to a modified international 10–20 system, and two reference electrodes were placed on both mastoid bones, as shown in Figure 4-A. The skin under the reference electrodes was abraded using Nuprep (Weaver and Co., USA) and disinfected with a 70% isopropyl alcohol swab before calibration. Of note, as shown in Figure 4-B, the impedance of the electrodes was calibrated to be under 5 kΩ using NaCl-based conductive gel (Quik-Gel, Neuromedical Supplies®). EEG signals from the electro-cap were amplified using the Scan NuAmps Express system (Compumedics Ltd., VIC, Australia) and recorded at a sampling rate of 500 Hz with 16-bit quantisation.

### EEG validation

Please note that all EEG data were saved in raw form, meaning that the pre-processing steps (filtering, baseline correction, artefact rejection) applied to the data during the experimental trials were not applied to the stored data. Before data analysis, we recommend that researchers pre-process the raw data using a digital bandpass filter (1–50 Hz) to remove line noise and artefacts and then down-sample to 250 Hz to reduce the data amount. Additionally, severe contamination of the EEG signals due to eye movement, blinking, muscle activity and environmental noise must be manually removed to minimise their effect on subsequent analysis.

Regarding data quality, the measured brain signals are easily contaminated with artefacts originating from non-cerebral origins. The amplitude of these artefacts, commonly generated by ocular and muscle activities, can be quite large and may mask the cortical signals of interest, biasing the analysis and interpretation. Furthermore, several blind source separation techniques have been proposed for signal pre-processing to remove such artefacts. For example, independent component analysis (ICA) produces the maximal temporally independent signals available in the EEG recording and is a powerful tool for suppressing artefacts.

Additionally, we shared this EEG dataset with our partner groups, including the University of California at San Diego (UCSD) and the DCS Corporation. Our findings are consistent with their results [20,21], which supports the technical validation of accurately estimating shifts in driver arousal, fatigue and vigilance levels by evaluating changes in behavioural and neurocognitive performance.

**Usage Notes**

The experimental data can be downloaded from figshare with publicly accessible repository (Data Citation 1). Any researcher interested in this dataset can sign up to figshare and download the project named "Multi-channel EEG recordings during a sustained-attention driving task" in user's personal computer.

The data can be analysed in EEGLAB, which is an interactive MATLAB toolbox with an interactive graphical user interface (GUI). It includes multiple functions for processing continuous and event-related EEG using ICA, time/frequency analysis and other methods including artefact rejection under multiple operation systems. EEGLAB has also provided extensive tutorials (https://sccn.ucsd.edu/wiki/EEGLAB_TUTORIAL_OUTLINE) to help researchers conduct data analysis. We recommend researchers use EEGLAB with version 5.03 on Windows 7 or Linux.

Of note, a data analysis tutorial (named "Tutorial Data Analysis for Multi-channel EEG Recordings during a Sustained-attention Driving Task.pdf) and MATLAB codes (named "Code-availability.zip") are regarded as references for EEG pre-processing and data analysis during a sustained-attention driving task. To access these items, please go to the URL: https://doi.org/10.6084/m9.figshare.6427334.v2. It can ensure that researchers can easily reuse the dataset.

Additionally, we provided some key notes for data analysis.
1. Load the existing dataset. Select menu item 'File' and select the 'Load existing dataset' sub-menu item. Then, a sub-window will pop up to select the existing dataset (e.g., s01_051017m.set).

2. Check the workspace in MATLAB. For the 'EEG' variable, some key information is explained below:
srate: sampling rate
EEG.chanlocs: the number of channels
EEG.event: event type and latency
data: EEG signals with channels multiply times

3. Extract data epochs and conduct further data analysis. To study the event-related EEG dynamics of continuously recorded data, we must extract the data epoch time of the events of interest (for example, the data epoch time of the onsets of one class of experimental stimuli) by selecting Tools>Extract Epochs. Additionally, removing a mean baseline value from each epoch is useful when there are baseline differences between data epochs (e.g., arising from low frequency drifts or artefacts). Additionally, EEGLAB contains several functions for plotting averages of dataset trials/epochs, selecting data epochs, comparing ERP images, working with ICA components, decomposing time/frequency information and combining multiple datasets.

## Acknowledgements
This work was supported in part by the Australian Research Council (ARC) under discovery grant DP180100670 and DP180100656. Research was also sponsored in part by the Army Research Laboratory and was accomplished under Cooperative Agreement Number W911NF-10-2-0022 and W911NF-10-D-0002/TO 0023. The views and the conclusions contained in this document are those of the authors and should not be interpreted as representing the official policies, either expressed or implied, of the Army Research Laboratory or the U.S Government. The U.S Government is authorized to reproduce and distribute reprints for Government purposes notwithstanding any copyright notation herein. Additionally, we express our gratitude to all of the subjects who kindly participated in this study. In addition, we thank all of the students and staff at the Brain Research Center in National Chiao Tung University and Computational Intelligence and Brain Computer Interface Center in University of Technology Sydney for their assistance during the study process.

## Author contributions
C.T. Lin and J.T. King designed and performed the experiment. Z. Cao and C.H Chuang collected, checked and analysed the data. Z. Cao and C.H. Chuang wrote the paper. C.T. Lin and J.T. King revised the manuscript. All of the authors read and approved the final manuscript.

## Competing interests
Z. Cao, C.H. Chuang, J.T. King and C.T. Lin have no competing financial interests to declare.

## Figure Legends
Figure 1. An event-related lane-departure paradigm in a virtual-reality (VR) dynamic driving simulator.

Figure 2. Experimental design. (A) Event-related lane-deviation paradigm. (B–C) EEG and behaviour were recorded simultaneously.

Figure 3. An example of behaviour and EEG performance. (A) Behaviour performance. (B) EEG signals with associated events.

Figure 4. The layout of electrodes and impedance of the EEG caps used in the experiments. (A) The blue electrodes use the international 10–20 system, and the green ones are additional electrodes on the cap. (B) The contact impedance between all of the electrodes and the skin was kept below 5 kΩ.

## Tables

| EEG.event.type | 251 | 252 | 253 | 254 |
|---|---|---|---|---|
| Definition | Deviation onset (left) | Deviation onset (right) | Response onset | Response offset |

## Data Citations

**Figures**

Figure 1

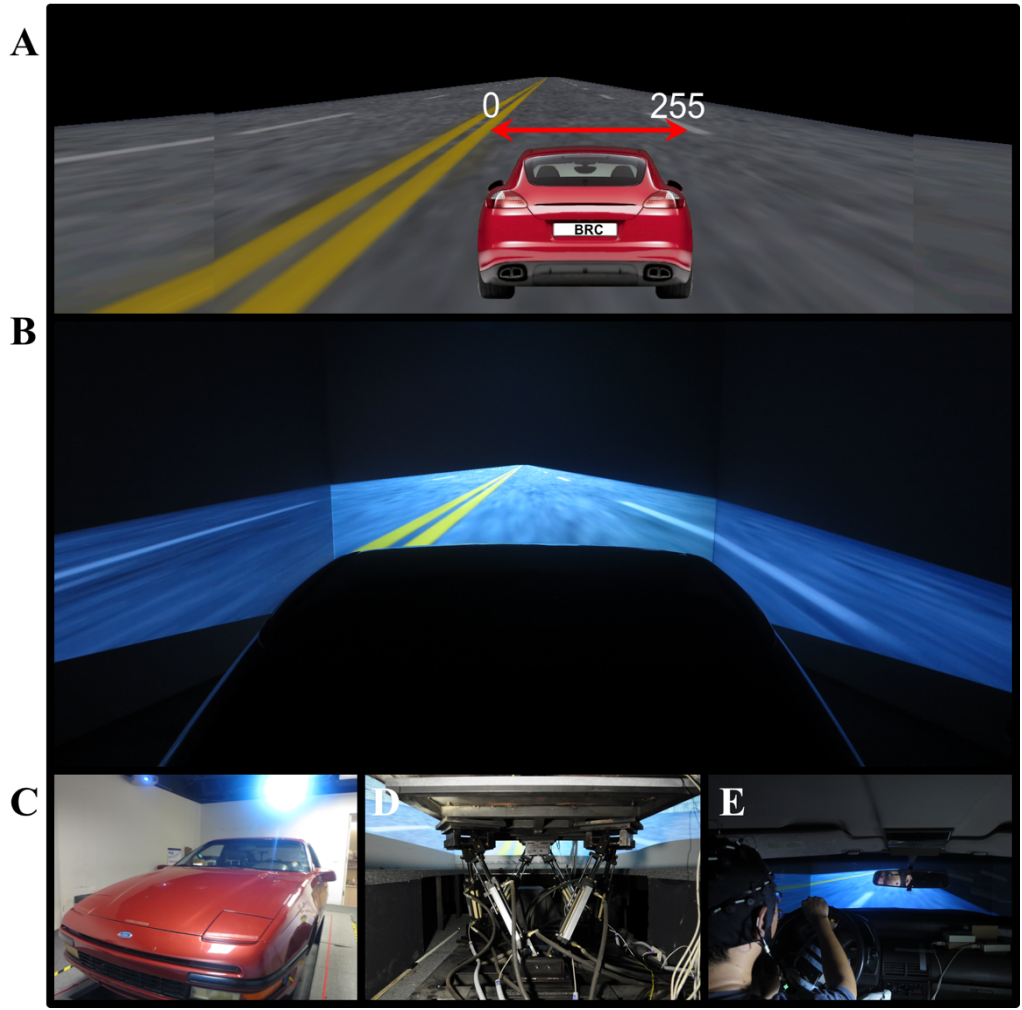

Figure 2

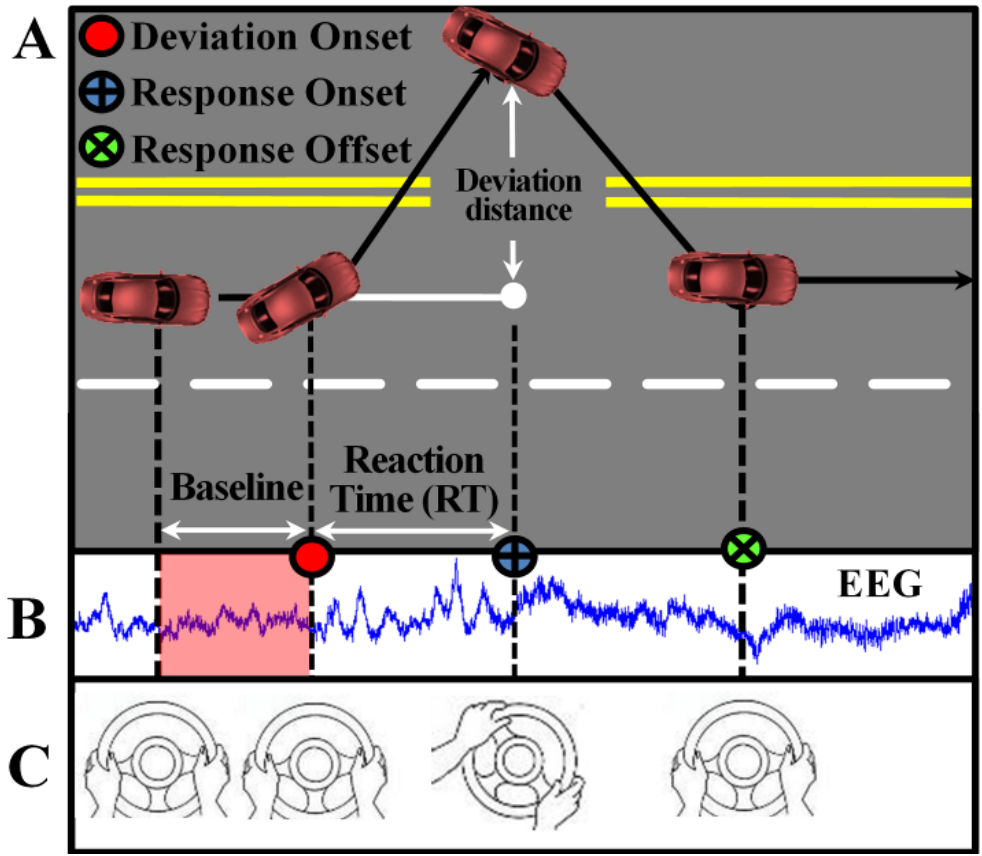

Figure 3

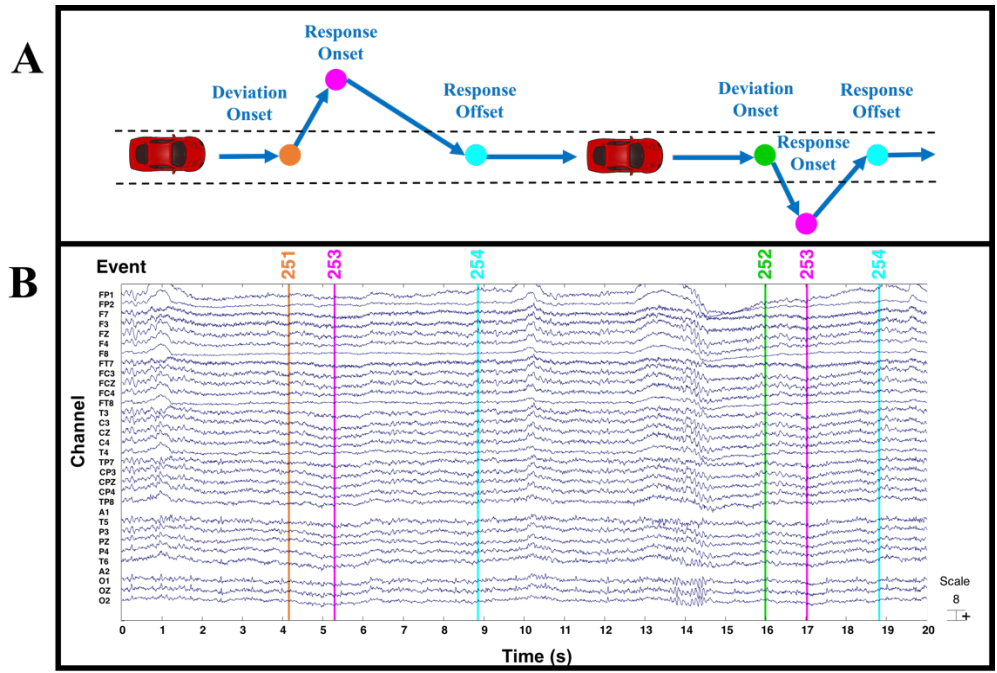

Figure 4

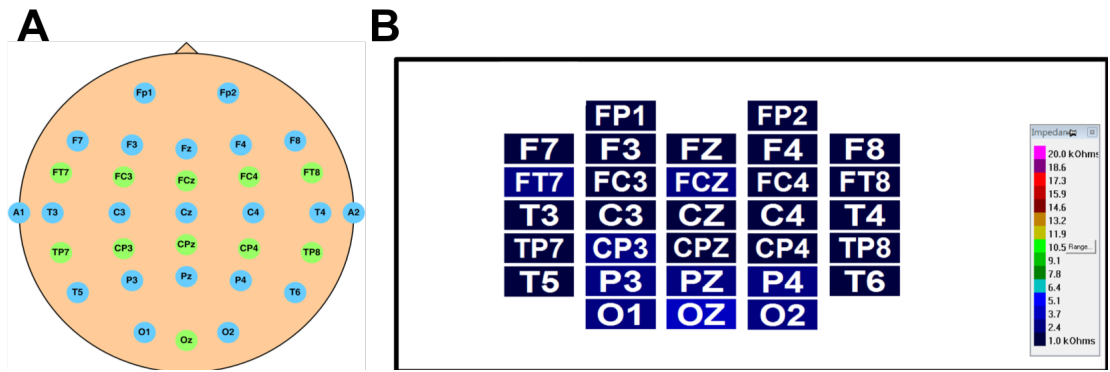